\documentclass{moriond}

\def\Journal#1#2#3#4{{#1} {\bf #2}, #3 (#4)}

\def\NPB{{\em Nucl. Phys.} B}
\def\PLB{{\em Phys. Lett.}  B}
\def\PRL{\em Phys. Rev. Lett.}

\def\be{\begin{equation}}
\def\ee{\end{equation}}
\def\bea{\begin{eqnarray}}
\def\eea{\end{eqnarray}}

\newcommand{\Photo}{\includegraphics[height=35mm]{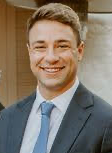}}

\begin{document}
\vspace*{4cm}
\title{Single top and rare top quark production (including FCNC searches) at ATLAS and CMS}

\author{Samuel May \\ on behalf of the ATLAS and CMS Collaborations}

\address{Boston University, Department of Physics \\ 
590 Commonwealth Avenue, Boston, MA 02215}

\maketitle\abstracts{
Measurements of single top and rare top quark production may be sensitive
to theories of physics beyond the standard model (SM), including those in which the energy scale of new physics is beyond the energies
directly accessible at the Large Hadron Collider.
Such models may be observable through signatures like forbidden SM interactions or deviations of the top quark's couplings from the SM predictions.
An overview of recent searches from the  ATLAS and CMS Collaborations is presented, with a focus on flavor-changing neutral currents.
}

\section{Introduction}
\footnote{Copyright 2022 CERN for the benefit of the ATLAS and CMS Collaborations. Reproduction of this article or parts of it is allowed as specified in the CC-BY-4.0 license.}
The standard model of particle physics (SM) is known to be incomplete; its inability to provide viable candidates for dark matter and dark energy and its inability to explain the observed matter-antimatter asymmetry are examples of its multiple shortcomings.
Despite strong motivation for physics beyond the standard model (BSM), searches at LHC experiments have not yet found conclusive evidence of its presence.
This lack of discovery motivates the possibility that BSM phenomena may be present at energies which are beyond those directly accessible at the LHC, as well as experimental searches for signatures of these theories.
The top quark may serve as a sensitive probe of such models, through signatures forbidden in the SM like flavor-changing neutral currents (FCNCs) and charged lepton flavor violation (CLFV) or through modifications of the top quark's couplings.

An overview of recent searches and measurements from the ATLAS~\cite{atlas} and CMS~\cite{cms} Collaborations is presented.
Searches for FCNCs are described in section~\ref{sec:fcnc},
CLFV in section~\ref{sec:clfv},
and measurements of single top quark production in association with a Z boson are summarized in section~\ref{sec:tzq}.
Unless noted otherwise, the searches presented use the full Run 2 dataset from the LHC, corresponding to integrated luminosities between 137 and 139 fb$^{-1}$.
Several of the presented analyses interpret results in the context of the effective field theory (EFT) framework, which parametrizes deviations from the SM in terms of coefficients of dimension-six operators added to the SM Lagrangian (SMEFT)~\cite{eft1,eft2}.

\section{Flavor-changing neutral currents}
\label{sec:fcnc}
\begin{figure}
\begin{minipage}{0.58\linewidth}
\centerline{\includegraphics[width=0.95\linewidth]{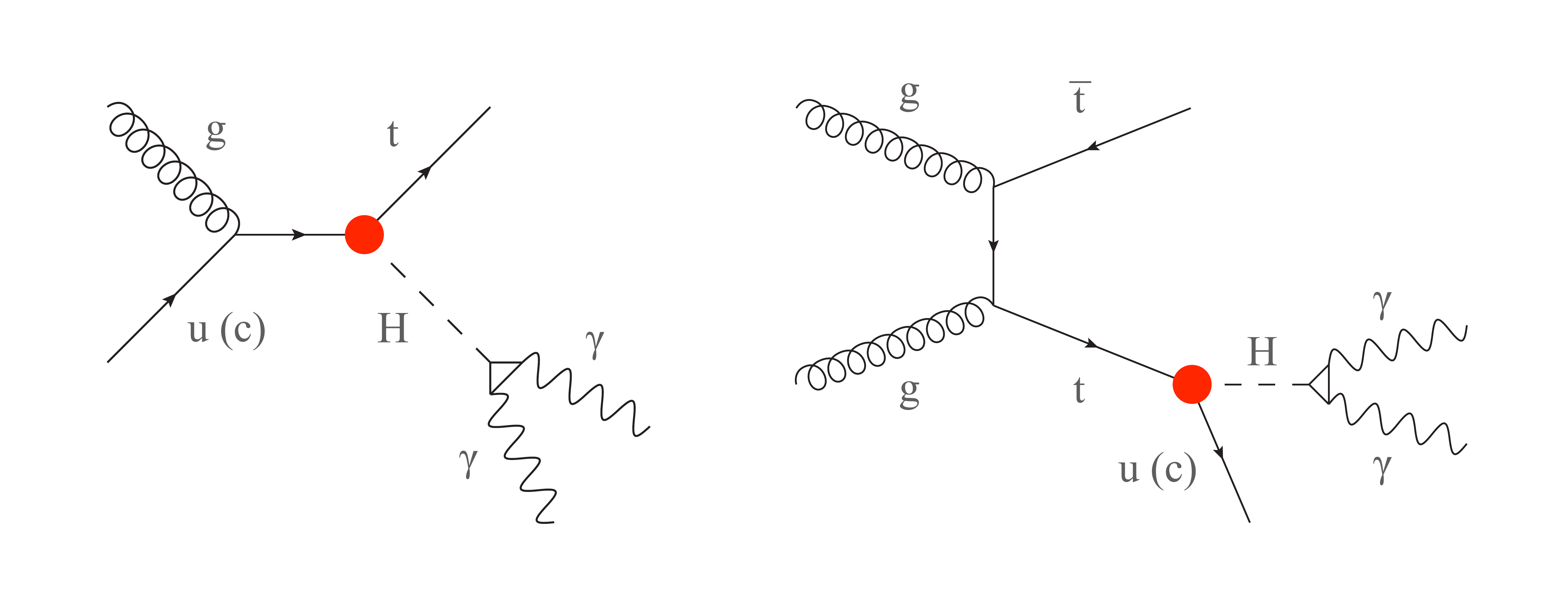}}
\end{minipage}
\hfill
\begin{minipage}{0.42\linewidth}
\centerline{\includegraphics[width=0.95\linewidth]{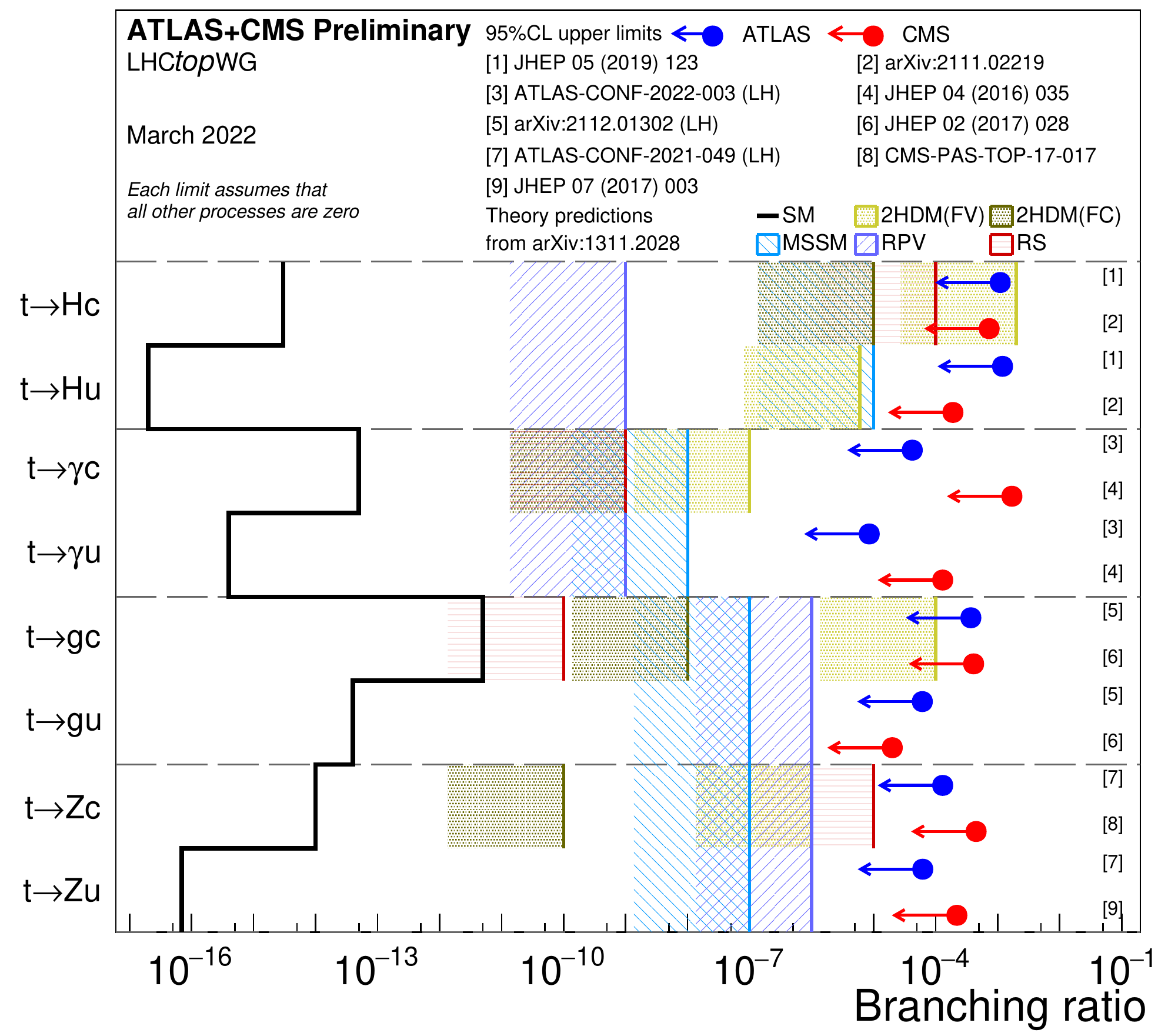}}
\end{minipage}
\caption[]{
Current experimental exclusion limits on the top quark FCNC branching fractions from ATLAS (blue) and CMS (red), with ranges of predicted branching fractions for various BSM models in shaded bands. Figure taken from Ref.~\cite{atlas_top_public_results}.}
\label{fig:fcnc}
\end{figure}

In the SM, flavor-changing quark decays mediated by netural currents are forbidden at tree level. While they are allowed in higher-order diagrams, their rates are suppressed by the Glashow–Iliopoulos–Maiani mechanism and Cabibbo–Kobayashi–Maskawa unitarity constraints, resulting in SM branching fractions which are less than $10^{-12}$~\cite{topwg}, well below current LHC sensitivity.
Any observation of FCNCs would then be an unambiguous sign of new physics.
A variety of BSM models predict enhanced rates of $\mathrm{t} \to \mathrm{Xq}$, where X may be any of the neutral bosons: photon ($\gamma$), gluon (g), Z boson (Z) or Higgs boson (H).
Searches for FCNC interactions often exploit two distinct production modes: the associated production of a single top quark with a neutral mediator via an up or charm quark (Fig.~\ref{fig:fcnc} left) and the decay of a top quark to a neutral mediator and an up or charm quark in $\mathrm{t}\bar{\mathrm{t}}$ production (Fig.~\ref{fig:fcnc} middle).
The current experimental limits on the top quark FCNC branching fractions are summarized in the right-panel of Fig.~\ref{fig:fcnc}.

\subsection{Searches for FCNCs with $t \to Hq$}

A search for $\mathrm{t} \to \mathrm{Hq}$ FCNC interactions with the Higgs boson decaying to a bottom quark-antiquark pair with is presented by the CMS Collaboration~\cite{fcnc_hbb}.
The search uses a partial Run 2 dataset corresponding to an integrated luminosity of 101 fb$^{-1}$.
Leptonic decays of the top quark are targeted by using single lepton (e and $\mu$) triggers to reduce the large multi-jet background.
A boosted decision tree (BDT)-based discriminant is used in a binned likelihood fit to extract a potential signal.
No excess above the background prediction is observed and the subsequent 95\% confidence level (CL) observed (expected) upper limits on $\mathcal B(\mathrm{t} \to \mathrm{Hu})$ are 0.079\% (0.110\%) and on $\mathcal B(\mathrm{t} \to \mathrm{Hc})$ are 0.094\% (0.086\%).

A search for $\mathrm{t} \to \mathrm{Hq}$ FCNC interactions in final states with two photons is presented by the CMS Collaboration~\cite{fcnc_hgg}.
Despite its small branching fraction of 0.2\%, the H decay to two photons benefits from small backgrounds and good resolution of the invariant mass of the diphoton system ($m_{\gamma \gamma}$), typically between 1 and 2 GeV.
An ensemble of BDTs is trained to distinguish the signal from two classes of backgrounds: SM H production modes (resonant backgrounds) and other SM processes (non-resonant backgrounds).
The data are consistent with the background-only hypothesis and the resulting 95\% CL upper limits on $\mathcal B(\mathrm{t} \to \mathrm{Hu})$ are 0.019\% (0.031\%) and on $\mathcal B(\mathrm{t} \to \mathrm{Hc})$ are 0.073\% (0.051\%), which are the most stringent experimental limits to-date. 

A search for $\mathrm{t} \to \mathrm{Hq}$ FCNC interactions in events with a pair of $\tau$-leptons is presented by the ATLAS Collaboration~\cite{fcnc_htt}.
At least one tau lepton is required to decay hadronically, and is identified with a deep neural network (DNN), while the other tau lepton may decay either hadronically or to an electron or a muon.
Events are categorized by the number of jets, number of leptons (e or $\mu$), and number of hadronic taus.
Discriminants based on the output of BDTs are then used to extract a potential signal.
A slight excess above the background prediction is observed in both the search for $\mathrm{t} \to \mathrm{Hu}$ and for $\mathrm{t} \to \mathrm{Hc}$, with a statistical significance of 2.3$\sigma$ in both cases.
The observed (expected) 95\% CL upper limits are 0.072\% (0.036\%) for $\mathcal B(\mathrm{t} \to \mathrm{Hu})$ and 0.099\% (0.050\%) for $\mathcal B(\mathrm{t} \to \mathrm{Hc})$, as shown in Fig.~\ref{fig:clfv_tzq}.

\subsection{Searches for FCNCs mediated by a gluon, Z boson, or photon}
A search for $\mathrm{t} \to \mathrm{gq}$ FCNC interactions is presented by the ATLAS Collaboration~\cite{fcnc_tgq}.
As gluon-initiated jets are produced in huge numbers in pp collisions, the analysis searches only for FCNC interactions mediated by a gluon through single top associated production.
Exactly one lepton (e or $\mu$), at least one hadronic jet, and missing transverse momentum are required in the final state.
The kinematics of the signal process vary depending on whether the quark in the initial state is a valence quark, the dominant production mechanism for $u + g \to t$, or a sea quark, the dominant production mechanism for $\mathrm{c} + \mathrm{g} \to \mathrm{t}$.
Two separate DNNs are trained for each case.
There is no observed excess above the background prediction and 95\% CL upper limits of $0.61 \times 10^{-4}$ and $3.7 \times 10^{-4}$ are set on $\mathcal B(\mathrm{t} \to \mathrm{gu})$ and $\mathcal B(\mathrm{t} \to \mathrm{gc})$, respectively.
In the EFT framework, limits on the Wilson coefficients of dimension-6 operators are $|C_{uG}^{\mathrm{ut}}| / \Lambda^2 < 0.057$ TeV$^{-2}$ and $|C_{uG}^{\mathrm{ct}}| / \Lambda^2 < 0.14$ TeV$^{-2}$.

A search for $\mathrm{t} \to \mathrm{Zq}$ FCNC interactions is presented by the ATLAS Collaboration~\cite{fcnc_tzq}.
Events are required to have three leptons (e or $\mu$), two of which must form an opposite-sign same-flavor pair with a mass within 15 GeV of $m_Z$.
An ensemble of BDTs is used to further separate signal from the SM backgrounds.
No excess is observed and 95\% CL upper limits on the branching ratios for $\mathcal B(\mathrm{t} \to \mathrm{Zu})$ ($\mathcal B(\mathrm{t} \to \mathrm{Zc})$) of $6.2 \times 10^{-5}$ ($13 \times 10^{-5}$) are set in the cases of a left-handed coupling and $6.6 \times 10^{-5}$ ($12 \times 10^{-5}$) in the case of a right-handed coupling.

A search for $\mathrm{t} \to \gamma \mathrm{q}$ FCNC interactions is presented by the ATLAS Collaboration~\cite{fcnc_tgammaq}.
The analysis utilizes a final state with exactly one photon, one lepton (e or $\mu$), one b-tagged hadronic jet, and missing transverse momentum.
A multiclass DNN is used to assign events to one of the signal production modes (single top FCNC production or $\mathrm{t}\bar{\mathrm{t}}$ FCNC decay) or background.
The data are consistent with the background-only hypothesis and the resulting 95\% CL upper limits on $\mathcal B(\mathrm{t} \to \gamma \mathrm{u})$ ($\mathcal B(\mathrm{t} \to \gamma \mathrm{c})$) are $0.85 \times 10^{-5}$ ($4.2 \times 10^{-5}$) in the cases of a left-handed coupling and $1.2 \times 10^{-5}$ ($4.5 \times 10^{-5}$) in the case of a right-handed coupling.

\section{Charged lepton flavor violation}
\label{sec:clfv}
A search for charged lepton flavor violation  in both single top quark associated production and $\mathrm{t}\bar{\mathrm{t}}$ decay is presented by the CMS Collaboration~\cite{clfv}.
Recent measurements of $B$ meson decays are in tension with SM predictions and may hint at violation of lepton universality.
Some models of new physics proposed to explain these tensions, such as certain leptoquark models, also imply charged lepton flavor violation with a branching fraction of $\mathrm{t} \to \mathrm{llc}$ (l = e or $\mu$) of $\mathcal O(10^{-6})$.
Events in this analysis are required to have an opposite-sign e/$\mu$ pair and at least one b-tagged jet.
The region with more than one b-tagged jet is used as a control region to model the dominant $\mathrm{t}\bar{\mathrm{t}}$ background, while the region with exactly one b-tagged jet serves as a signal region.
A BDT is trained with the missing transverse momentum and lepton \& jet kinematics to distinguish signal from $\mathrm{t}\bar{\mathrm{t}}$. 
No significant excess is found and 95\% CL upper limits on $\mathcal B(\mathrm{t} \to \mathrm{e}\mu \mathrm{u})$ ($\mathcal B(\mathrm{t} \to \mathrm{e}\mu \mathrm{c})$) are found to be $0.13 \times 10^{-6}$ ($1.31 \times 10^{-6}$), $0.07 \times 10^{-6}$ ($0.89 \times 10^{-6}$), and $0.25 \times 10^{-6}$ ($2.59 \times 10^{-6}$) in the cases of vector, scalar, and tensor CLFV interactions, respectively.

\begin{figure}
\begin{minipage}{0.52\linewidth}
\centerline{\includegraphics[width=0.95\linewidth]{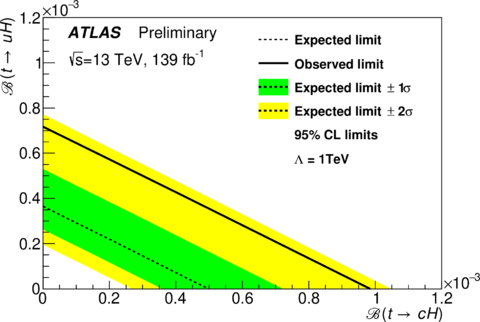}}
\end{minipage}
\hfill
\begin{minipage}{0.36\linewidth}
\centerline{\includegraphics[width=0.95\linewidth]{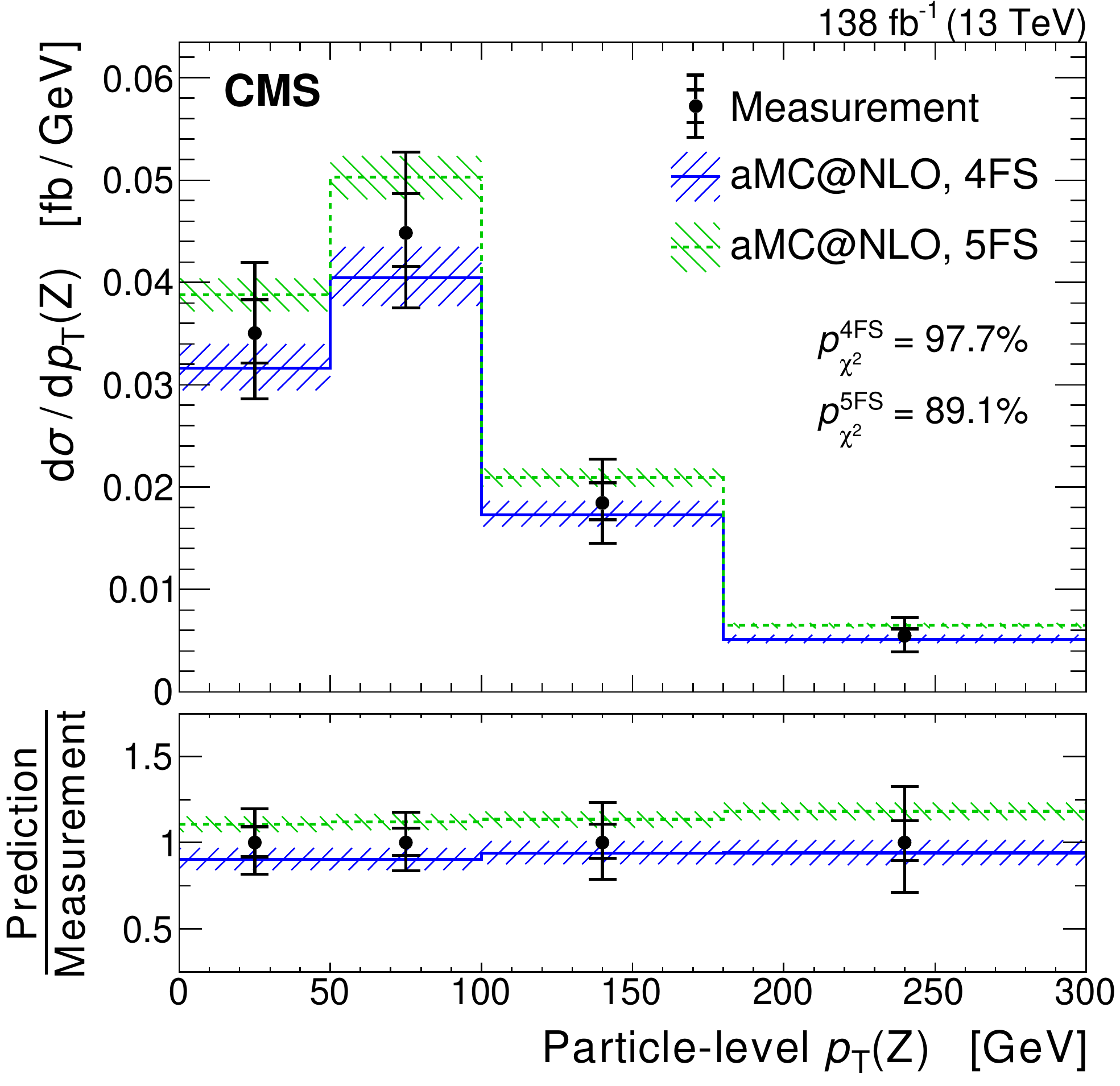}}
\end{minipage}
\caption[]{
Observed and expected limits on $\mathcal B(\mathrm{t} \to \mathrm{Hc})$ vs. $\mathcal B(\mathrm{t} \to \mathrm{Hu})$ in the $\mathrm{H} \to \tau \tau$ final state~\cite{fcnc_htt} (left),
and differential cross sections as a function of $p_{T}(Z)$ at the particle-level~\cite{tzq} (right).
}
\label{fig:clfv_tzq}
\end{figure}

\section{Measurements of single top quark production in association with a Z boson}
\label{sec:tzq}
The first differential measurement and the most precise inclusive measurement to-date of tZq production in final states with three leptons are presented by the CMS Collaboration~\cite{tzq}.
Two of the three leptons must form an opposite-sign same-flavor pair with an invariant mass within 15 GeV of $m_Z$.
Events are then categorized by the multiplicities of jets and b-tagged jets and the outputs of multivariate classifiers are used to extract the signal strength.
The ratio of the observed cross section to the SM prediction is $\mu_{\mathrm{tZq}} = 0.933^{+0.080}_{-0.077}$ (stat)$^{+0.078}_{-0.064}$ (syst) and is the most precise determination to-date.
Differential measurements, like that of the transverse momentum of the Z boson shown in Fig.~\ref{fig:clfv_tzq}, are performed at both parton- and particle-level and found to be in agreement with the SM predictions in both the four- and five-flavor schemes.

\section{Summary}
Searches for signatures of new physics in single and rare top quark production from the ATLAS and CMS collaborations are presented,
including those for flavor-changing neutral currents, charged lepton flavor violation, and measurements of the tZq process.
Results are so far consistent with the SM predictions, further restricting the allowed ranges for a variety of BSM theories, as shown in Fig.~\ref{fig:fcnc}.
While no evidence of new physics is found, the larger datasets expected from Run 3 and the HL-LHC will help to further clarify the landscape.

\end{document}